\documentclass[11pt,a4paper]{article}
\pdfoutput=1
\usepackage{graphicx}
\usepackage{verbatim}
\usepackage{epsfig}
\usepackage{amsmath}
\usepackage{amssymb}
\usepackage{times}
\usepackage[ruled]{algorithm2e}

\DeclareGraphicsExtensions{.pdf}
\graphicspath{{figures/}}

\date{}

\begin{document}

\title{Evolution of Cooperation and Coordination in a Dynamically Networked Society}
\author{%
E. Pestelacci\footnote{Information Systems Department, University of Lausanne, Switzerland} \and 
M. Tomassini%
\addtocounter{footnote}{-1}%
\footnotemark
\and L. Luthi%
\addtocounter{footnote}{-1}%
\footnotemark
}
\maketitle

\vspace{-0.4cm}

\begin{abstract}
\noindent Situations of conflict giving rise to social dilemmas are widespread in society and game theory is one
major way in which they can be investigated. Starting from the observation that individuals in society
interact through networks of acquaintances, we model the co-evolution of the agents' strategies and
of the social network itself using two prototypical games, the Prisoner's Dilemma and the Stag Hunt. 
Allowing agents to dismiss ties and establish new ones, we find that 
cooperation and coordination can be achieved through the self-organization of the social network,
a result that is non-trivial, especially in the Prisoner's Dilemma case. The evolution and stability of cooperation implies the condensation of agents exploiting particular game strategies
into strong and stable clusters which are more densely connected, even in the more difficult case
of the Prisoner's Dilemma.
\end{abstract}

\section{Introduction}
\label{intro} 

In this paper we study the behavior of a population of agents playing some simple
two-person, one-shot non-cooperative game.
Game theory~\cite{myerson} deals with social interactions where two or more individuals take
decisions that will mutually influence each other. It is thus a view of collective systems
in which global social outcomes emerge as a result of the interaction of the individual
decisions made by each agent. Some extremely simple games lead to puzzles and dilemmas that
have a deep social meaning. The most widely known among these games is the Prisoner's Dilemma (PD), a universal metaphor for the tension
that exists between social welfare and individual selfishness.
It stipulates that, in situations where individuals may either cooperate or defect,
they will rationally choose the latter.
However, cooperation would be the preferred outcome when global welfare is considered.
Other simple games that give rise to social dilemmas are the Hawk-Dove and the Stag-Hunt (SH) games.

In practice, however, cooperation and coordination on common objectives is often seen in human and animal
societies ~\cite{axe84,skyrms}. Coordinated behavior, such as having both players cooperating in the SH, is a bit less
problematic as this outcome, being a Nash equilibrium, is not ruled out by theory. For the PD, in which cooperation is
theoretically doomed between rational agents, several mechanisms have been invoked to explain the emergence of cooperative behavior.
Among them, repeated interaction, reputation, and belonging to a recognizable group have often been mentioned \cite{axe84}.
Yet, the work of Nowak and May~\cite{nowakmay92} showed that the simple fact that players are arranged according
to a spatial structure and only interact with neighbors is sufficient to sustain a certain amount of cooperation even when the game is played anonymously and without repetition.
Nowak and May's study and much of the following work were based on regular structures such as two-dimensional grids (see~\cite{nowak-sig-00} for a recent review). 
Nevertheless, many actual social networks usually have a topological structure that is neither regular nor random but rather of the \textit{small-world} type.
Roughly speaking, small-world networks are graphs in which any node is relatively close to any other node.
In this sense, they are similar to random graphs but unlike regular lattices.
However, in contrast with random graphs, they also have a certain amount of local structure,
as measured, for instance, by a quantity called the \textit{clustering coefficient} which
essentially represents the probability that two neighbors of a given node are themselves connected (an excellent
review of the subject appears in~\cite{newman-03}). 
Some work has been done in recent years in the direction of using those more realistic kind
of networks, including actual social networks. In particular we mention
Santos and Pacheco's work on scale-free networks~\cite{santos-pach-05}, work on Watts--Strogatz small-world graphs~\cite{social-pd-kup-01,tom-luth-giac-06}, and 
on model and real social networks~\cite{luthi-pest-tom-physa07}.  A recent contribution focuses on
repeated games and learning~\cite{csermely-07} and Szab\'o and F\'ath have 
published an excellent and very complete review of work done up to 2006~\cite{Szabo-Fath-07}. 
These investigations have convincingly shown that a realistic structure of the society, with interactions mainly limited to neighbors in the network, is well sufficient in allowing cooperative and coordinated behavior to emerge without making any particular assumption about the rationality of the actors or their computational and forecasting capabilities.

Most of the above mentioned studies have assumed a fixed population size and structure, which amounts to dealing with a closed system and ignoring any fluctuations in the system's size and internal interactions. 
However, real social networks, such as friendship or collaboration networks, are not in an equilibrium state, but are open systems that continually evolve with new agents joining or leaving the network, and relationships (i.e.~links in network terms) being made or dismissed
by agents already in the network~\cite{barab-collab-02,kossi-watts-06,tom-leslie-evol-net-07}.
Thus, the motivation of the present work is to re-introduce these coupled dynamics into our model and to investigate under which conditions, if any, cooperative and coordinated behavior may emerge and be stable. 
In this paper, we shall deal with networked populations in which
the number of players remains constant but the interaction structure,~i.e. who interacts with whom, does not stay fixed; on the contrary, it changes in time and
its variation is dictated by the very games that are being played by the agents.
A related goal of the present work is to study the topological structures of the emergent networks and their relationships with the strategic choices of the agents.\\
Some previous work has been done on evolutionary games on dynamic 
networks~\cite{lut-giac-tom-al-06,santos-plos-06,skyrms-pem-00,zimmer-pre-05}. Skyrms and Pemantle~\cite{skyrms-pem-00}
was recently brought to our attention by a reviewer. It is one of the first important attempts to study the kind of networks that
form under a given game and, as such, is closely related to the work we describe here.  The main ideas are
similar to ours: agents start interacting at random according to some game's payoff matrix and, as they evolve their
game strategy according to their observed payoffs, they also have a chance of breaking ties and forming
new ones, thus giving rise to a social network. The main differences with the present work is that the number
of agents used is low, of the order of $10$ instead of the $10^3$ used here. This allows us to study the 
topological and statistical nature of the evolving networks in a way that is not possible with a few agents, while
Skyrms' and Pemantle's work is more quantitative in the study of the effects of the stochastic dynamics on
the strategy and network evolution process. The work of Zimmermann and Egu\'iluz~\cite{zimmer-pre-05} 
is based on similar considerations too. There is a rather large population which has initially a random
structure. Agents in the population play the one-shot two-person Prisoner's Dilemma game against each 
other and change their strategy by copying the strategy of the more successful agent in their neighborhood.
They also have the possibility of dismissing interactions between defectors and of rewiring them randomly
in the population. The main differences with the present work are the following. Instead of just considering
symmetrical undirected links, we have a concept of two directed, weighted links between  pairs of agents. In our
model there is  a finite probability
of breaking any link, not only links between defectors, although defector-defector and cooperator-defector
links are much more likely to be dismissed than cooperator-cooperator links. When a link is broken it
is rewired randomly in~\cite{zimmer-pre-05} while we use a link redirection process which favors neighbors
with respect to more relationally distant agents. In~\cite{zimmer-pre-05} only the Prisoner's Dilemma is studied
and using a reduced parameter space. We study both the Prisoner's Dilemma and the Stag Hunt games
covering a much larger parameter space. Concerning timing of events, we use an asynchronous update policy for the agents'
strategies, while update is synchronous in~\cite{zimmer-pre-05}. Finally, instead of a best-takes-over discrete rule, we use a
smoother strategy update rule which changes an agent's strategy with a probability proportional to the payoffs difference. Santos et al.~\cite{santos-plos-06} is a more recent paper also dealing with similar issues. However,
they use a different algorithm for severing an undirected link between two agents which, again, does not
include the concept of a link weight. Furthermore, the Stag Hunt game is only mentioned in passing, and their
strategy update rule is different. In particular, they do not analyze in detail the statistical structure of the emerging
networks, as we do here. Other differences with the above mentioned related works will be described in the
discussion and analysis of results. Finally, our own previous work~\cite{lut-giac-tom-al-06} also deals with
the co-evolution of strategy and structure in an initially random network. However, it is very different from
the one presented here since we used a semi-rational threshold decision rule for a family of games similar,
but not identical to the Prisoner's Dilemma in~\cite{lut-giac-tom-al-06}. Furthermore,
the idea of a bidirectional weighted link between agents was absent, and link rewiring was random.

This article is structured as follows. 
In sect.~\ref{social_dilemmas}, we give a brief description of the games used in our study. This part is intended to make the article self-contained. 
In sect.~\ref{model}, we present a detailed description of our model of co-evolving dynamical networks. 
In sect.~\ref{simul}, we present and discuss the simulation results and their significance for the social networks. 
Finally, in sect.~\ref{conclusions}, we give our conclusions and discuss possible extensions and future work.

\section{Social Dilemmas}
\label{social_dilemmas}

The two representative games studied here are the Prisoner's Dilemma (PD) and the
Stag-Hunt (SH) of which we briefly summarize the significance and the main results. More detailed accounts
can be found elsewhere, for instance in \cite{axe84,skyrms}.
In their simplest form, they are  two-person, two-strategies, symmetric
games with the following payoff bi-matrix:

\begin{table}[hbt]
\begin{center}
{\normalsize
\begin{tabular}{c|cc}
 & C & D\\
\hline
C & (R,R) & (S,T)\\
D & (T,S) & (P,P)
\end{tabular}
}
\end{center}
\vspace{-0.5cm}
\end{table}

\noindent In this matrix, R stands for the \textit{reward}
the two players receive if they
both cooperate (C), P is the \textit{punishment} for bilateral defection (D), and T  is the
\textit{temptation}, i.e. the payoff that a player receives if it defects, while the
other cooperates. In this case, the cooperator gets the \textit{sucker's payoff} S.
In both games, the condition $2R > T + S$ is imposed so that
mutual cooperation is preferred over an equal probability of unilateral cooperation and defection.
For the PD, the payoff values are ordered numerically in the following way: $T > R > P > S$. 
Defection is always the best rational individual choice in the PD; 
(D,D) is the unique \textit{Nash equilibrium} (NE) and also an \textit{evolutionarily stable strategy} (ESS) \cite{myerson,weibull95}.
Mutual cooperation  would be preferable but it is a strongly dominated strategy.

In the SH, the ordering is $R > T > P > S$, which means that mutual cooperation (C,C) is the best outcome,
Pareto-superior, and
a Nash equilibrium. However, there is a second equilibrium in which both players defect
(D,D) and which is somewhat ``inferior'' to the previous one, although perfectly equivalent from
a NE point of view. The (D,D) equilibrium is less satisfactory yet  ``risk-dominant'' since 
playing it ``safe'' by choosing strategy D guarantees at least a payoff of P, while
playing C might expose a player to a D response by her opponent, with the ensuing minimum payoff S.
Here the dilemma is represented by the fact that the
socially preferable coordinated equilibrium (C,C) might be missed for ``fear'' that the other player
will play D instead. 
There is a third
mixed-strategy NE in the game, but it is commonly dismissed because of its
inefficiency and also because it is not an ESS \cite{weibull95}.
Although the PD has received much more attention in the literature than the SH, the latter is also
very useful, especially as a metaphor of coordinated social behavior for mutual benefit. These aspects
are nicely explained in~\cite{skyrms}.

\section{Model Description}
\label{model}

Our model is strictly local as no player uses information other than the one concerning the player itself and 
the players it is directly connected to. In particular, each agent knows its own current strategy and payoff,
and the current strategies and payoffs of its immediate neighbors. Moreover, as the model is an evolutionary one, no rationality, in the 
sense of
game theory, is needed~\cite{weibull95}. Players just adapt their behavior such that they copy more successful strategies in their environment  with higher probability, a process commonly called \textit{imitation} in the
literature~\cite{hofb-sigm-book-98}. Furthermore, they are able to locally
assess the worth of an interaction and possibly dismiss a relationship that does not pay off enough.
The model and its dynamics are described in detail in the following sections.

\subsection{Network and Interaction Structure}
\label{net}
The network of agents will be represented as an undirected graph $G(V,E)$, where the
 set of vertices $V$ represents the agents, while the set of edges (or links) $E$ represents their symmetric interactions. The
 population size $N$ is the cardinality of $V$. A neighbor  of an agent $i$ is any other agent $j$ such that there is an edge $\{ij\}  \in E$.
The set of neighbors of $i$  is called  $V_i$  and its cardinality is the degree $k_i$ of vertex $i \in V$. The average
degree of the network will be called $\bar k$. 

\begin{figure} [!ht]
\begin{center}
	\vspace*{-0.2cm}
\includegraphics[width=4.5cm] {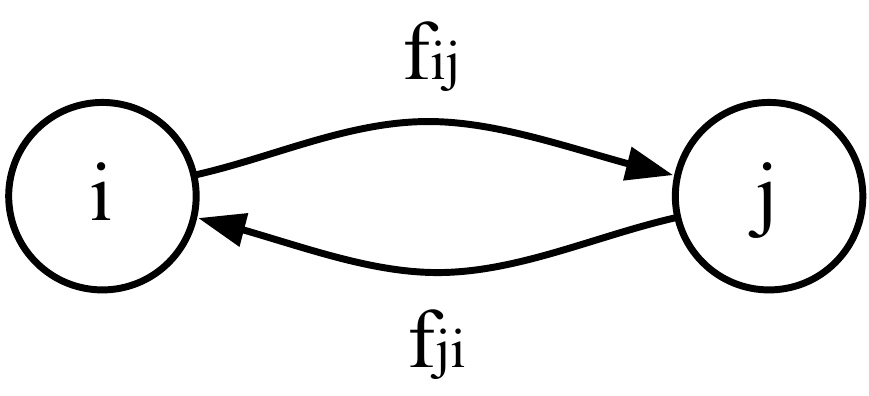} \protect \\
\vspace{-0.2cm}	
\caption{Schematic representation of mutual trust between two agents through the strengths
of their links.\label{force}}
\end{center}
\end{figure}

Although from the network structure point of view there is a single undirected link between a player $i$ and another player $j \in V_i$, we shall maintain
two links: one going from $i$ to $j$ and another one in the reverse direction (see fig.~\ref{force}). Each
link has a weight or ``force'' $f_{ij}$ (respectively $f_{ji}$). This weight, say $f_{ij}$,  represents in an indirect way an
abstract quality that could be related to the
``trust'' player $i$ attributes to player $j$, it may take any value in $[0,1]$ and its
variation is dictated by the payoff earned by $i$ in each encounter with $j$, as explained below. 

We point out that we do not believe that this model could represent, however roughly, a situation of
genetic relatedness in a human or animal society. In this case, at the very least, one should have at the
outset that link strengths between close relatives should be higher than the average forces in the
whole network and such groups should form cliques of completely connected agents. In contrast, we start
our simulations from random relationships and a constant average link strength (see below). Thus, our
simplified model is closer to one in which relationships between agents are only of socio-economic
nature.

The idea behind the introduction of the forces $f_{ij}$ is loosely inspired by the potentiation/depotentiation of
connections between neural networks, an effect known as the \textit{Hebb rule} \cite{hebb}. In our context, it can be seen
as a kind of  ``memory'' of previous encounters. However, it must be distinguished from the memory used in
 iterated games, in which
players ``remember'' a certain amount of previous moves and can thus conform their future strategy on the
analysis of those past encounters~\cite{myerson}. Our interactions are strictly one-shot, i.e. players ``forget'' the results of
previous rounds and cannot recognize previous partners and their possible playing patterns. However, a certain amount of past history is implicitly
contained in the numbers $f_{ij}$ and this information may be used by an agent when it will come to decide
whether or not an interaction should be dismissed (see below)\footnote{ A further refinement of the concept could
take obsolescence phenomena into account. For instance, in the same way that pheromone trails
laid down by ants evaporate with time, we could introduce a progressive loss of strength of the links
proportional to the time during which there is no interaction between the concerned agents.
For the sake of simplicity, we prefer to stick with the basic model in this work}.
This bilateral view of a relationship is, to our knowledge, new in evolutionary game models
on graphs.

We also define a quantity $s_i$ called \textit{satisfaction} of an agent $i$ which is the sum of all the weights
of the links between $i$ and its neighbors $V_i$ divided by the total number of links $k_i$:

$$ s_i = \frac{\sum_{j \in V_i} f_{ij} } {k_i}. $$

\noindent We clearly have $0 \le s_i \le 1$. 

\subsection{Initialization}
\label{init}

The constant size of the network during the simulations is $N=1000$. The initial graph is generated randomly with
a mean degree comprised between $\bar k=5$ and $\bar k=20$. These values of $\bar k$ are of the order of those actually found in many social networks (see, for
instance,~\cite{barab-collab-02,kossi-watts-06,newman-collab-01-1,TLGL-GPEM-07}).
Players are distributed uniformly at random over the graph vertices with 50\% cooperators. Forces
 between any pair of neighboring players are initialized at $0.5$.
With $\bar k > 1$ a random graph finds itself past the percolation phase transition~\cite{bollobas} and thus it
has a giant connected component of size $O(N)$ while all the other components are of size $O(log(N))$.
We do not assure that the whole graph is connected, as isolated nodes will draw a random link during the dynamics (see below).

Before starting the simulations, there is another parameter $q$ that has to be set. This is akin to a
``temperature'' or noise level; $q$ is a real number in $[0,1]$ and
it represents the frequency with which an agent wishes to dismiss a link with one of its neighbors. The higher
$q$, the faster the link reorganization in the network. This parameter has a role analogous to the ``plasticity'' of
~\cite{zimmer-pre-05} and it controls the speed at which topological changes occur in the network. 
As social networks may structurally evolve at widely different speeds, depending
on the kind of interaction between agents, this factor might play a role in the model. For example, e-mail networks change their structure at a faster pace 
than, say, scientific collaboration networks~\cite{kossi-watts-06,tom-leslie-evol-net-07}. A similar coupling of time scales between strategy
update and topological update also occurs in~\cite{skyrms-pem-00,santos-plos-06}.

\subsection{Timing of Events}
\label{timing}

Usually, agents systems such as the present one, are updated synchronously, especially in evolutionary
game theory 
simulations~\cite{luthi-pest-tom-physa07,nowakmay92,santos-pach-05,zimmer-pre-05}. 
However, there are doubts about the physical signification of simultaneous
update~\cite{hubglance93}. For one thing, it is strictly speaking physically unfeasible as it would require a global clock,
while real extended systems in biology and society in general have to take into account finite signal propagation speed.
Furthermore, simultaneity may cause some artificial effects in the dynamics which are not observed
in real systems~\cite{hubglance93,lut-giac-tom-al-06}. Fully asynchronous update, i.e.~updating a randomly chosen agent at a time with or without
replacement also seems a rather arbitrary extreme case that is not likely to represent reality very accurately. In view of these
considerations, we have chosen to update our population in a partially synchronous manner. In practice, we
define a fraction $f =n/N$ (with $N = an, a \in \mathbb{N} $) and, at each simulated discrete time step, we update only $n  \le N$ agents randomly
chosen with replacement. This is called a \textit{microstep}. After $N/n$ microsteps, called a \textit{macrostep}, $N$ agents will have been
updated, i.e. the whole population will have been updated in the average. With $n = N$ we recover the fully synchronous update, while
$n=1$ gives the extreme case of the fully asynchronous update. Varying $f$ thus allows one to investigate the role
of the update policy on the dynamics. We study several different values of $f$, but we mainly focus on $f=0.01$. 

\subsection{Strategy and Link Dynamics}
\label{strat-link-dyn}

Here we describe in detail how individual strategies, links, and link weights are updated. Once a given
node $i$ is chosen to be activated, i.e. it belongs to the fraction $f$ of nodes that are to be updated
in a given microstep, $i$ goes through the following steps:

\begin{itemize}
\item if the degree of agent $i$, $k_i = 0$ then
 player $i$ is an isolated node. In this case a link with strength $0.5$ is created
from $i$ to a player $j$ chosen uniformly at random among the other $N-1$ players in the network. 
\item otherwise,
\begin{itemize}
\item either  agent $i$ updates its strategy according to a local \textit{replicator dynamics} rule with probability $1-q$ or, with probability $q$, agent $i$ may delete a link with a given neighbor $j$ and creates a new $0.5$ force link with another node $k$ ;

\item the forces between $i$ and its neighbors $V_i$ are updated 

\end{itemize}
\end{itemize}

Let us now describe each step in more detail.

\paragraph{Strategy Evolution.}

We use a local version of replicator dynamics (RD) as described in~\cite{hauer-doeb-2004}  and further
modified in~\cite{luthi-pest-tom-physa07} to take into account the fact that the number of neighbors in a degree-inhomogeneous
network can be different for different agents. 
The local dynamics of a player $i$ only depends on its own strategy and on the strategies of the $k_i$ players in its neighborhood $V_i$.
Let us call $\pi_{ij}$ the payoff player $i$ receives when interacting with neighbor $j$. This payoff is defined as
$$
\pi_{ij} =  \sigma_i(t)\; M\; \sigma_{j}^T(t),
$$
\noindent where $M$ is the payoff matrix of the game (see sect.~\ref{social_dilemmas}) and $\sigma_i(t)$ and $\sigma_j(t)$ are the strategies played by $i$ and $j$ at time $t$.
The quantity
$$
 \widehat{\Pi}_i(t) = \sum _{j \in V_i}\pi_{ij}(t)
$$
\noindent is the \textit{accumulated payoff} collected by player $i$ at time step $t$.
The rule according to which agents update their strategies is the conventional RD in which strategies that do
better than the average increase their share in the population, while those that fare worse than average
decrease.
To update the strategy of player $i$, another player $j$ is drawn at random from the neighborhood $V_i$.
It is assumed that the probability of switching strategy is a function $\phi$ of the payoff difference, where 
$\phi$ is a monotonically increasing function~\cite{hofb-sigm-book-98}. 
Strategy $\sigma_i$ is replaced by $\sigma_j$ with probability
$$
 p_i = \phi(\widehat{ \Pi}_j - \widehat{\Pi}_i).
$$ 
The major differences with standard RD is that two-person encounters between players are only possible among neighbors, instead of being drawn from the whole population, and the latter is finite in our case.
Other commonly used strategy update rules include imitating the best in the 
neighborhood~\cite{nowakmay92,zimmer-pre-05}, or replicating in proportion to the payoff ~\cite{hauer-doeb-2004,tom-luth-giac-06}.
Although, these rules are acceptable alternatives, they do not lead to replicator dynamics and will not be dealt with here. 
We note also that the straight accumulated payoff $\widehat{\Pi}_i$ has a technical problem when used on degree-inhomogeneous
systems such as those studied here, where agents (i.e.~nodes) in the network may have different numbers of
neighbors. In fact, in this case $\widehat{\Pi}_i$  does not induce invariance of the RD with respect to affine 
transformations of the game's payoff matrix as it should~\cite{weibull95}, and makes the results depend
on the particular payoff values. 
Thus, we shall use a modified accumulated payoff $\Pi$ instead as defined in~\cite{luthi-pest-tom-physa07}. This payoff, which is the standard accumulated payoff corrected with a factor that
takes into account the variable number of neighbors an agent may have,  does not suffer from the standard accumulated payoff limitations.

\paragraph{Link Evolution.}

The active agent $i$, which has $k_i \ne 0$ neighbors will, with probability $q$, attempt to dismiss an interaction with one of its neighbors. This
is done in the following way.
Player $i$ will look at its satisfaction $s_i$. The higher $s_i$, the more satisfied the player, since a high satisfaction
is a consequence of successful strategic interactions
with the neighbors. Thus, there should be a natural tendency to try to dismiss a link when $s_i$ is low. This is simulated by drawing a uniform pseudo-random number $r \in [0,1]$ and breaking a link when $r \ge s_i$.
Assuming that the decision is taken to cut a link, which one, among the possible $k_i$, should be chosen?
Our solution again relies on the strength of the relevant links. First a neighbor $j$ is chosen with probability proportional to $1-f_{ij}$, i.e. the stronger the link, the less likely it will be chosen. This intuitively corresponds to $i$'s observation that it is preferable to dismiss an interaction with a neighbor $j$ that has contributed little to $i$'s payoff over several rounds of play. However, in our system dismissing a link is not free: $j$ may
``object'' to the decision. The intuitive idea is that, in real social situations, it is seldom possible to take unilateral
decisions: often there is a cost associated, and we represent this hidden cost by a probability 
$1 - (f_{ij} + f_{ji})/2$
with which $j$ may refuse to be cut away. In other words, the link is less likely to be deleted if $j$ appreciates $i$, i.e. when $f_{ji}$ is high.  A simpler solution would be to try to cut the weakest link, which is what happens most
of the time anyway. However, with a finite probability of cutting any link, our model introduces a small
amount of noise in the process which can be considered like ``trembles'' or errors in game theory~\cite{myerson}
 and which
roughly reproduces decisions under uncertainty in the real world.

\begin{figure} [!ht]
\begin{center}
	\vspace*{-0.2cm}
\includegraphics[width=7cm] {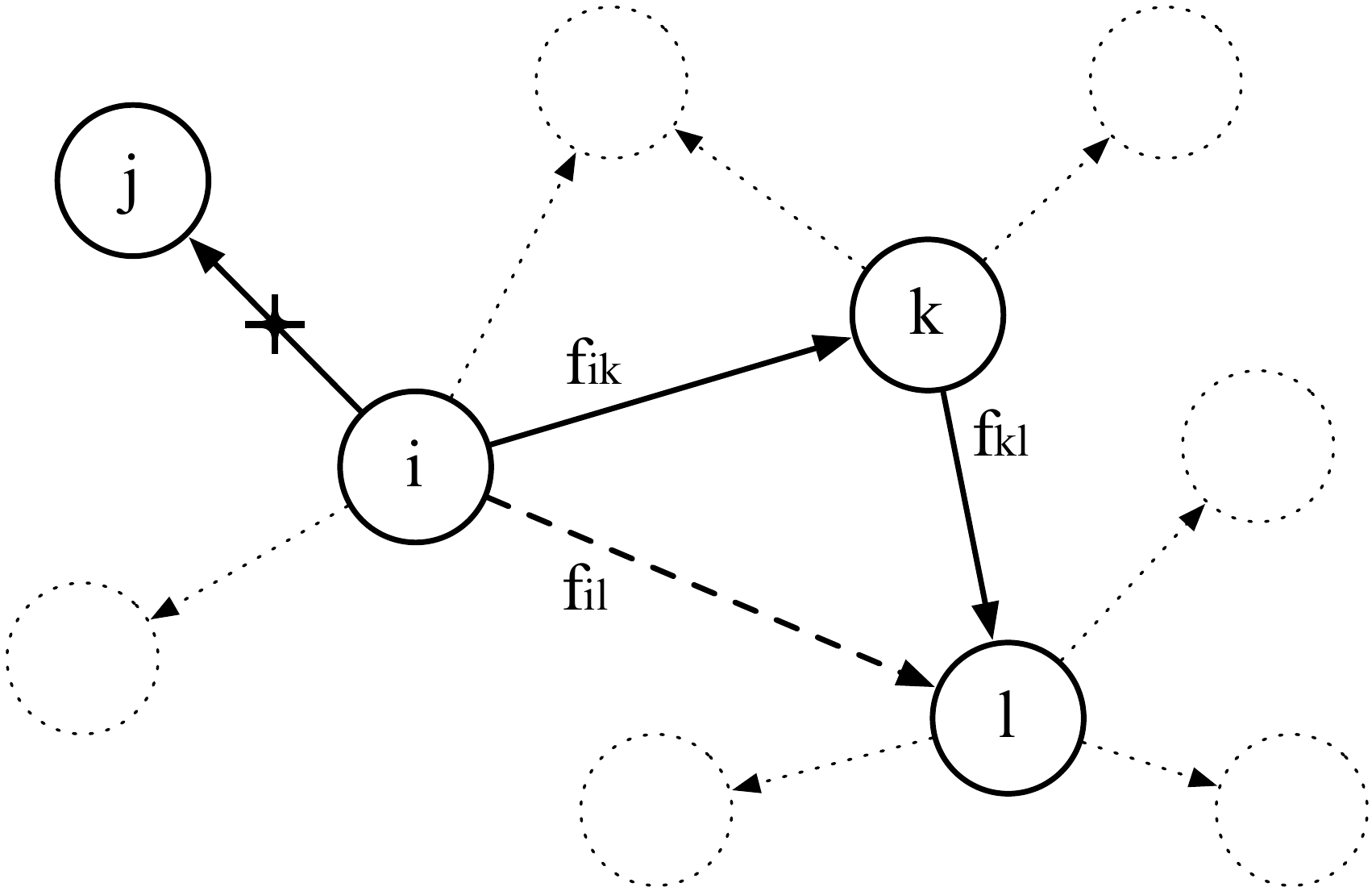} \protect \\
	\vspace*{-0.2cm}
\caption{Illustration of the rewiring of link $\{ij\}$ to $\{il\}$. Agent $k$ is
chosen to introduce player $l$ to $i$ (see text). \label{rewire}}
\end{center}
\end{figure}

Assuming that the $\{ij\}$ link is finally cut, how is a new link to be formed? 
The solution adopted here is inspired by the observation that, in social networks, links are
usually created more easily between people who have a mutual acquaintance than those who do not.
First, a neighbor $k$ is chosen in $V_i \setminus \{j\}$ with probability proportional to $f_{ik}$, thus favoring neighbors $i$ trusts.
Next, $k$ in turn chooses player $l$ in his neighborhood $V_k$ using the same principle, i.e. with probability proportional to $f_{kl}$. If $i$ and $l$ are not connected,
a link $\{il\}$ is created, otherwise the process is repeated in $V_l$. Again, if the selected node, say $m$, is not connected to $i$, a new link $\{im\}$ is established. If this also fails, a new link between $i$ and a randomly chosen node is created. 
In all cases the new link is initialized with a strength of $0.5$ in both directions.
This rewiring process is schematically depicted in fig.~\ref{rewire} for the case in which a link can be successfully established between players $i$ and $l$ thanks to their mutual acquaintance $k$.

At this point, we would like to stress several important differences with previous work in which links can be
dismissed in evolutionary games on networks~\cite{lut-giac-tom-al-06,santos-plos-06,zimmer-pre-05}. In~\cite{zimmer-pre-05}, only links between defectors are allowed to be cut unilaterally
and the study is restricted to the PD. Instead,
in our case, any link has a finite probability to be abandoned, even a profitable link between cooperators if it is
recent, although
links that are more stable, i.e. have high strengths, are less likely to be rewired. This smoother situation is made possible thanks to
our bilateral view of a link which is completely different from the undirected choice made
 in ~\cite{zimmer-pre-05}.

In~\cite{santos-plos-06}, links can  be cut by an unsatisfied player, where the concept of satisfaction is different from ours, and simply means that a cooperator or a defector will wish to break a link with a defector. 
The cut will be done with a certain probability that depends on the strategies of the two agents involved and their respective payoffs.
Once a link between $i$ and $j$ is actually cut and, among the two players, $i$ is the one selected to
 maintain the link, the link is rewired to a random neighbor of $j$.
If both $i$ and $j$ wish to cease their interaction, the link is attributed to $i$ or $j$ probabilistically, as a function of the respective payoffs of $i$ and $j$, and rewiring takes place from there.
Thus, although both $i$'s and $j$'s payoffs are taken into consideration in the latter case, there is no analogous of our ``negotiation'' process as the concept of link strength is absent. 
In~\cite{lut-giac-tom-al-06} links are cut according to a threshold decision rule and are rewired randomly anywhere in the network.

A final observation concerns the evolution of $\bar k$ in the network. While in~\cite{santos-plos-06,zimmer-pre-05} the initial mean degree is strictly maintained during network
evolution through the rewiring process, here it may increase slightly owing to the existence of isolated agents which, when chosen to be updated, will create a new link with another random agent.
While this effect is of minor importance and only causes small fluctuations of $\bar k$, we point out that in real evolving networks the mean connectivity fluctuates too~\cite{barab-collab-02,kossi-watts-06,tom-leslie-evol-net-07}.

\paragraph{Updating the Link Strengths.}

Once the chosen agents have gone through their strategy or link update steps, the strengths of the
links are updated accordingly in the following way:

$$ f_{ij}(t+1) = f_{ij}(t) + \frac {\pi_{ij} - \bar\pi_{ij}}  {k_i(\pi_{max} - \pi_{min}) },    $$

\noindent where $\pi_{ij}$ is the payoff of $i$ when interacting with $j$, $\bar\pi_{ij}$ is the payoff earned by
$i$ playing with $j$, if $j$ were to play his other strategy, and $\pi_{max}$ ($\pi_{min}$) is the maximal (minimal) possible payoff obtainable in a single interaction. This update is performed in both
directions, i.e. both $f_{ij}$ and $f_{ji}$ are updated $\forall j \in V_i$ because
both $i$ and $j$ get a payoff out of their encounter.\\

\section{Simulation Results}
\label{simul}

\subsection{Simulation Parameters}
\label{sim-par}

We simulate on our networks the two games previously mentioned in sect.~\ref{social_dilemmas}.
For each game, we can explore the entire game space by limiting our study to the variation
of only two parameters per game. This is possible without loss of generality owing to the invariance of
Nash equilibria and replicator dynamics under positive affine transformations of the payoff matrix using our payoff scheme \cite{weibull95}.
In the case of the PD, we  set $R=1$ and $S=0$,
and vary $1 \leq T \leq 2$ and $0 \leq P \leq 1$.
For the SH, we decided to fix $R = 1$ and $S = 0$ and vary $0 \leq T \leq 1$ and $0 \leq P \leq T$.
The reason we choose to set $T$ and $S$ in both the PD and the SH is
to simply provide natural bounds on the values to explore of  the remaining two parameters.
In the PD case, $P$ is limited between $R=1$ and $S=0$
in order to respect the ordering of the payoffs ($T>R>P>S$) and $T$'s upper bound is equal to 2  due to the
$2R > T+S$ constraint.
Had we  fixed $R=1$ and $P=0$ instead, $T$ could be as big as desired,
provided $S \leq 0$ is small enough.
In the SH, setting $R=1$ and $S=0$ determines the range of $T$ and $P$ (since this time $R>T>P>S$).
Note however, that for this game the only valid value pairs of $(T,P)$ are those that satisfy the $T > P$ constraint.

As stated in sect.~\ref{init}, we used networks of size $N=1000$, randomly generated with an average degree $\bar k \in \{5,10,20\}$ and randomly initialized with 50\% cooperators and 50\% defectors.
In all cases, the parameters are varied between their two bounds in steps of 0.1.
For each set of values, we carry out 50 runs of at most 20000 macrosteps each, using
a fresh graph realization in each run. A run is stopped when all agents are using the same strategy, in order to 
be able to measure statistics for the population and for the structural parameters of the graphs. The system is considered to have reached
a pseudo-equilibrium strategy state when the strategy of the agents (C or D) does not change over 150 further
macrosteps, which means $15 \times 10^4$ individual updates. We speak of pseudo-equilibria  or steady states and not
of true evolutionary equilibria because, as we shall see below, the system never quite reaches a totally stable
state in the dynamical systems sense in our simulations but only transient states that persist for a
long time.

\subsection{Cooperation and Stability}
\label{coop}
\begin{figure} [!ht]
\begin{center}
	\vspace*{-0.2cm}
\includegraphics[width=14cm] {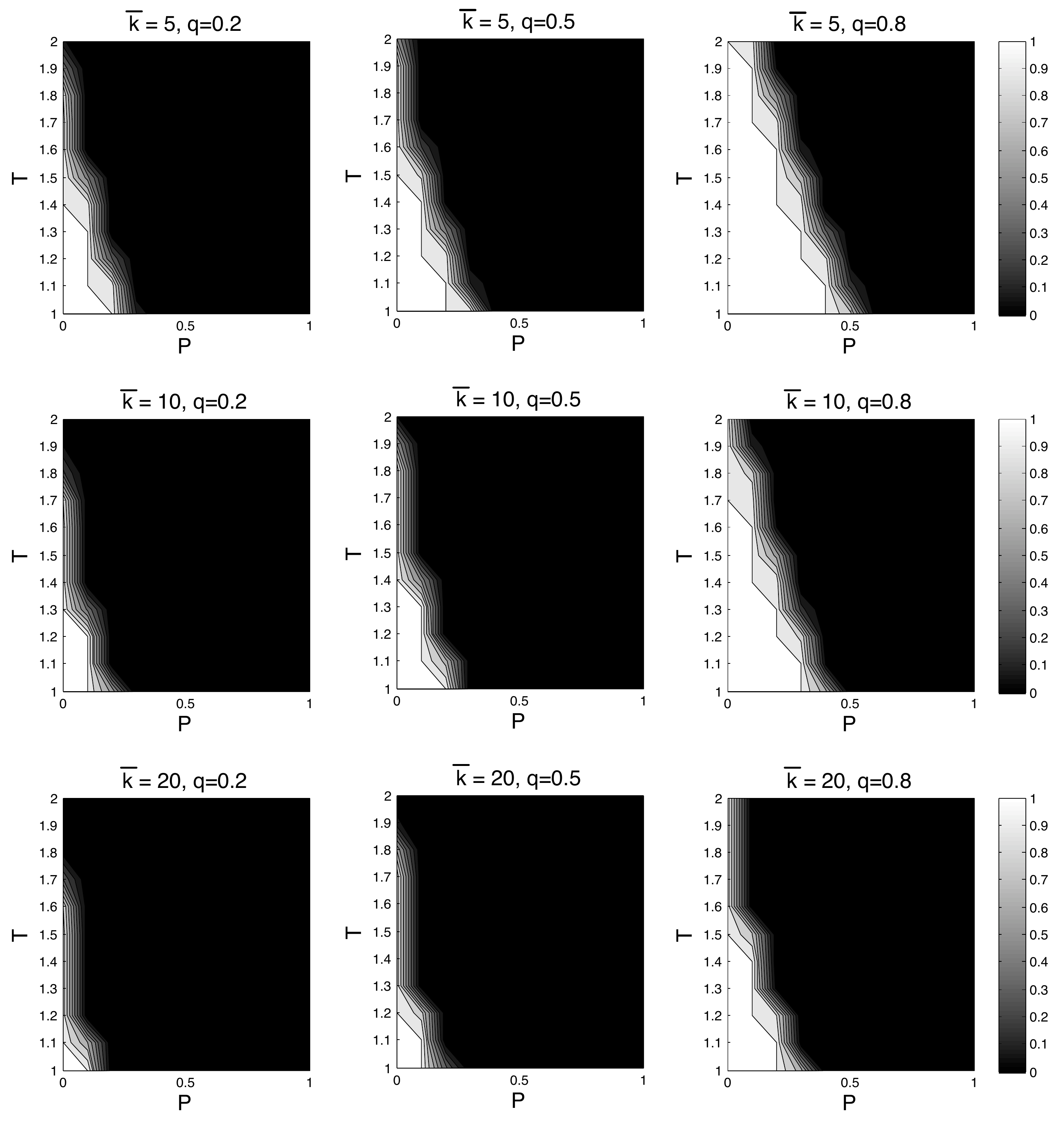} \protect \\	
\vspace*{-0.2cm}
\caption{Cooperation level for the PD in the game's configuration space. Darker gray means more
defection.\label{pd_coop}}
\end{center}
\end{figure}
\begin{figure} [!ht]
\begin{center}
	\vspace*{-0.2cm}
\includegraphics[width=15.5cm] {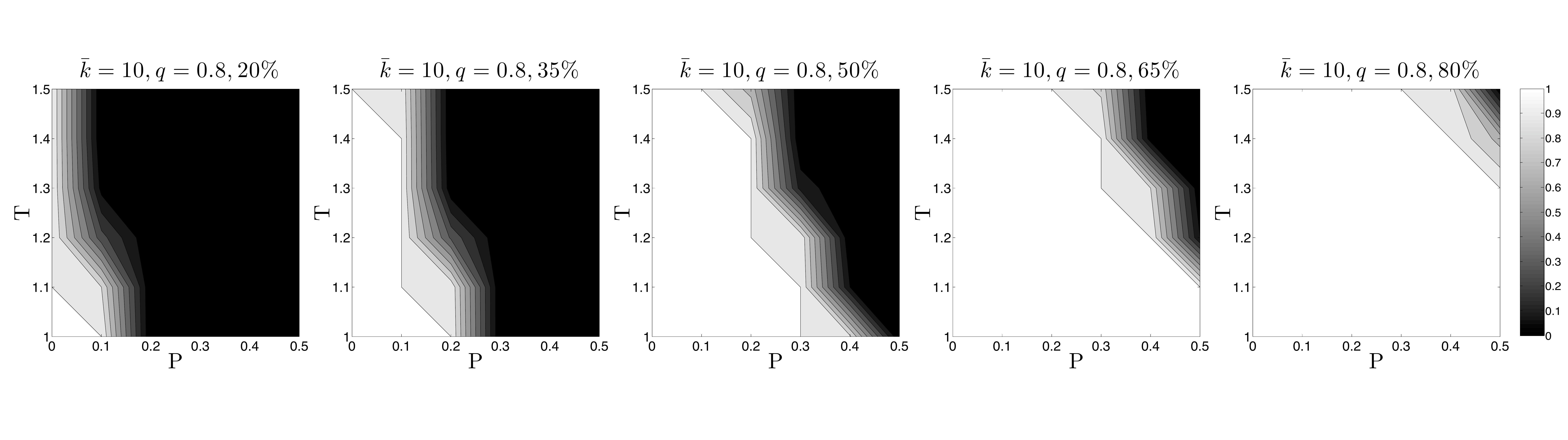} \protect \\	
\vspace*{-0.2cm}
\caption{Cooperation level for the PD starting with different fractions of cooperators increasing from $20\%$ to
$80\%$ from left to right. Only
the lower left quarter of the parameter space is shown. Results are the average of 50 independent runs.
\label{snapshots}}
\end{center}
\end{figure}

Cooperation results for the PD in contour plot form are shown in fig.~\ref{pd_coop}. We remark that,
as observed in other structured populations, cooperation may thrive in a small but non-negligible part of the parameter
space. Thus, the added degree of freedom represented by the possibility of refusing a partner and choosing
a new one does indeed help to find player's arrangements that help cooperation. This finding is in line
with the results of~\cite{santos-plos-06,zimmer-pre-05}. Furthermore, the fact that our artificial society 
model differs from the latter two in several important
ways also shows that the result is a rather robust one. When considering the dependence on the fluidity
parameter $q$, one sees in fig.~\ref{pd_coop} that the higher $q$, the higher the cooperation level. This was expected since
being able to break ties more often clearly gives cooperators  more possibilities for finding and keeping
fellow cooperators to interact with. This effect has been previously observed also in the works 
of~\cite{santos-plos-06,zimmer-pre-05} and, as such, seems to be a robust finding, relatively independent
of the other details of the models. The third para\-meter considered in fig.~\ref{pd_coop} is the mean degree $\bar k$.
For a given value of $q$, cooperation becomes weaker as $\bar k$ increases. We believe that, as far as $\bar k$ is concerned, a realistic average characterization
of actual social networks is represented by $\bar k=10$ (middle row in fig.~\ref{pd_coop}) as seen, for instance,
in~\cite{barab-collab-02,kossi-watts-06,newman-collab-01-1,TLGL-GPEM-07}. Higher average degrees do
exist, but they are found either in web-based pseudo-social networks or in fairly special collaboration networks
like the particle physics community, where it is customary to include as coauthors tens or even hundreds
of authors~\cite{newman-collab-01-1}. Clearly, there is a limit to the number of real acquaintances a given agent
may manage with. \\
We have also performed many simulations starting from different proportions of randomly
distributed cooperators and defectors to investigate the effect of this parameter on the evolution of
cooperation. In Fig.~\ref{snapshots} we show five different cases, the central image corresponding to
the $50\%$ situation. The images correspond to the lower left quarter of the right image in the middle row
 of Fig.~\ref{pd_coop}
with $\bar k =10$, $q=0.8$, $1<T<1.5$,  and $0 <  P < 0.5$.

Compared with the level of cooperation observed in simulations in
static networks, we can say that results are consistently better for co-evolving networks.  For example, the typical cases with
$\bar k=10$ and $q=0.5,0.8$ show significantly more cooperation than what was found in model and real social networks
in previous work~\cite{luthi-pest-tom-physa07}. Even when there is a much lower rewiring frequency, i.e.~with
$q=0.2$, the cooperation levels are approximately as good as those observed in our previous study in which exactly
the same replicator dynamics scheme was used to update the agents' strategies and the networks were of
comparable size. The reason for this behavior is to be found in the added constraints imposed by the
invariant network structure.
The seemingly contradictory fact that an even higher cooperation level may be reached in static
scale-free networks~\cite{santos-pach-05}, is theoretically interesting but easily dismissed as those graphs 
are unlikely models for social networks, which often
show fat-tailed degree distribution functions but not pure power-laws (see, for instance,~\cite{am-scala-etc-2000,newman-collab-01-1}). As a further indication of the latter, we shall see in sect.~\ref{topo} that, indeed,
emerging networks do not have a power-law degree distribution.

From the point of view of the evolutionary dynamics, it is interesting to point out that any given simulation run
either ends up in full cooperation or full defection. When the full cooperation state of the population is attained,
there is no way to switch back to defection by the intrinsic agent dynamics. In fact, all players are satisfied and have
strong links with their cooperating neighbors. Even though a small amount of noise may still be present when
deciding whether or not to rewire a link, since there are only cooperators around to imitate, there can be no
strategy change and only very little link rewiring. On the other hand, well before this stable state is reached and
there are still many defectors around, the system may experience some random drift that may drive it to full defection.
The converse may also happen, but
 when the full defection state is reached, the situation is qualitatively different. In this case
agents are unsatisfied, they will often try to rewire their links. However, all the other
players around being also defectors, there will be constant changes of the local network structure. Thus
the system will find itself in a fluctuating state, but this matters little for the bulk statistical properties of the
population and of the network. To be assured that this is indeed the case, we have conducted some very long
runs with all-defect end states. Global statistics do not change, except that the mean degree tends to increase slightly with time and
the degree distribution function continues to evolve (see sect.~\ref{topo}).
\begin{figure} [!ht]
\begin{center}
	\vspace*{-0.2cm}
\includegraphics[width=14cm] {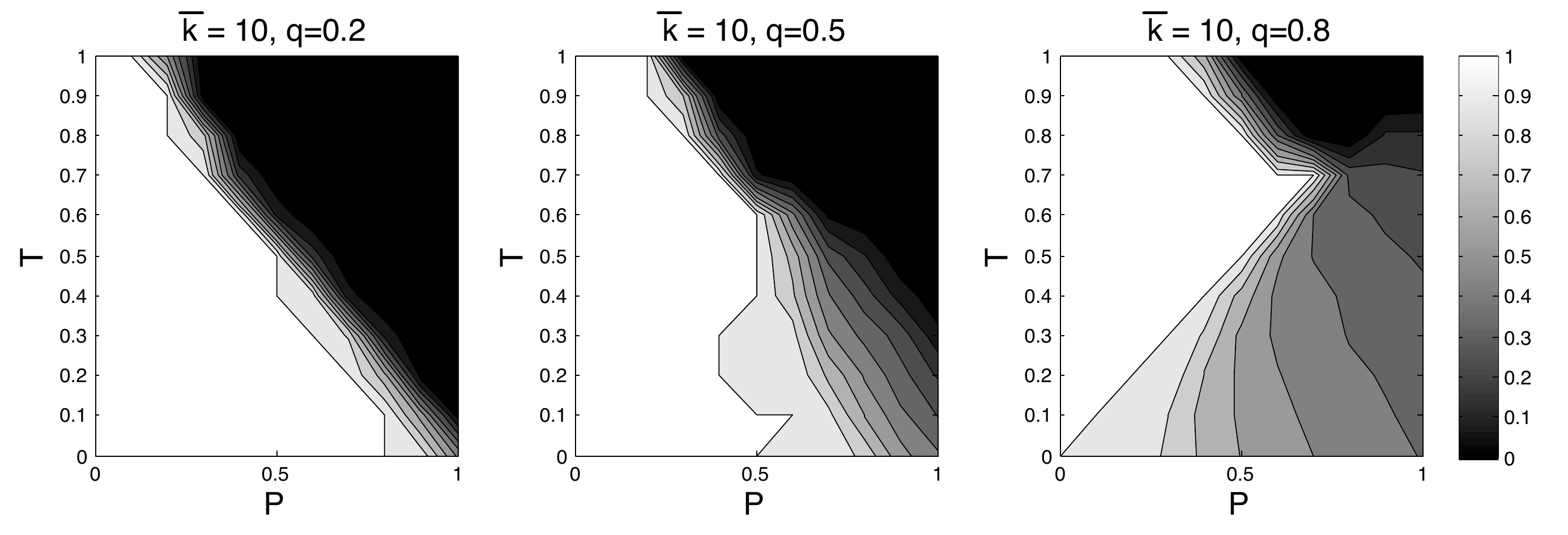} \protect \\	
	\vspace*{-0.2cm}
\caption{Cooperation level for the SH game.\label{sh_coop}}
\end{center}
\end{figure}

Cooperation percentages as a function of the payoff matrix parameters for the SH game are shown in fig.~\ref{sh_coop}
for $\bar k=10$ and $q=0.2,0.5$, and $0.8$.
Note that in this case only the upper left triangle of the configuration space is meaningful (see sect.~\ref{sim-par}).
The SH is different from the PD since there are two evolutionarily stable strategies which are therefore also
NEs: one population state in which everybody defects and the opposite one in which everybody cooperates
(see sect.~\ref{social_dilemmas}). Therefore, it is expected, and absolutely normal, that some runs will end up
with all defect, while others will witness the emergence of full cooperation. In contrast, in the PD the only theoretically
stable state is all-defect and cooperating states may emerge and be stable only by exploiting the graph
structure and creating more favorable neighborhoods by breaking and forming ties. The value of the SH is
in making manifest the tension that exists between the socially desirable state of full cooperation and the
socially inferior but less risky state of defection~\cite{skyrms}.  The final outcome of a given simulation run depends on the size of the basin of attraction
of either state, which is in turn a function of the relative values of the payoff matrix entries. To appreciate the
usefulness of making and breaking ties in this game we can compare our results with what is prescribed by the standard RD solution. Referring to the payoff table of sect.~\ref{social_dilemmas}, let's assume that the column player plays C with
probability $\alpha$ and D with probability $1-\alpha$. In this case, the expected payoffs of the row player
are:
$$
E_r[C] =\alpha R + (1-\alpha)S$$
and
$$E_r[D] =\alpha T + (1-\alpha)P
$$

The row player is indifferent to the choice of $\alpha$ when $E_r[C] = E_r[D]$. Solving for $\alpha$ gives:

\begin{equation}
 \alpha = \frac{P-S}{R-S-T+P}.
 \label{alpha}
\end{equation}

\begin{figure} [!ht]
\begin{center}
	\vspace*{-0.2cm}
\includegraphics[width=6cm] {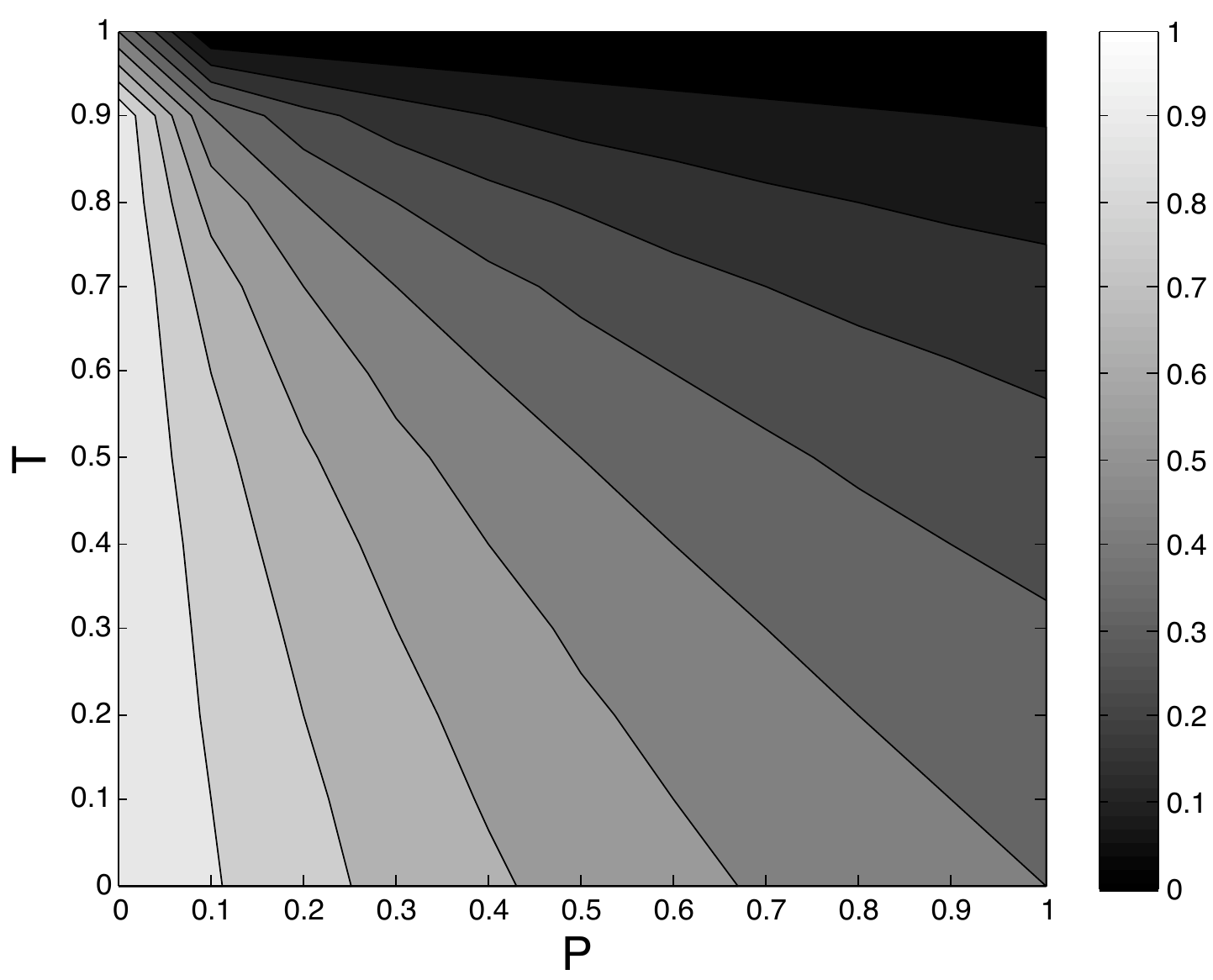} \protect \\	
	\vspace*{-0.2cm}
\caption{Probabilities of cooperation for the mixed strategy NE as a function of the
game's parameters for the Stag Hunt.\label{sh_theory}}
\end{center}
\end{figure}

Since the game is symmetric, the result for the column player is the same and $(\alpha C, (1-\alpha) D)$
is a NE in mixed strategies.
We have numerically solved the equation for all the sampled points in the game's parameter space, which
gives the results shown in fig.~\ref{sh_theory}. Let us now use the following payoff values 
 in order to bring them within the explored game space (remember that NEs
are invariant w.r.t. such a transformation~\cite{weibull95}):

\begin{table}[hbt]
\begin{center}
{\normalsize
\begin{tabular}{c|cc}
 & C & D\\
\hline
C & ($1,1$) & ($0,2/3$)\\
D & ($2/3,0$) & ($1/3,1/3$)
\end{tabular}
}
\end{center}
\vspace{-0.5cm}
\end{table}

Substituting in(\ref{alpha}) gives $\alpha=1/2$, i.e. the (unstable) polymorphic population should be
composed by about half cooperators and half defectors. Now, if one looks at fig.~\ref{sh_coop} at the
points where $P=1/3$ and $T=2/3$, one can see that this is approximately the case for the first 
image, within the limits of the approximations caused by the finite population size, the symmetry-breaking
caused by the non-homogeneous graph structure, and the local nature of the RD. On the other hand,
in the middle image and, to a greater extent,
in the rightmost image, this point in the game space corresponds to pure cooperation. In other words,
the non-homogeneity of the network and an increased level of tie rewiring has allowed the cooperation
basin to be enhanced with respect to the theoretical predictions of standard RD. Skyrms and Pemantle
found the same qualitative result for very small populations of agents when both topology and strategy
updates are allowed~\cite{skyrms-pem-00}. It is reassuring that coordination on the payoff-dominant 
equilibrium can still be achieved in large populations as seen here.

\subsection{Structure of the Emerging Networks}
\label{topo}

In this section we present a statistical analysis of the global and local properties of the networks that
emerge when the pseudo-equilibrium states of the dynamics are attained. Let us start by considering the
evolution of the average degree $\bar k$. Although there is nothing in our model to prevent a change
in the initial mean degree, the steady-state average connectivity tends to increase only slightly. For example,
in the PD with $q=0.8$ and $\bar k_{init}=5$ and $\bar k_{init}=10$, the average steady-state (ss) values are
$\bar k_{ss} \simeq 7$ and $\bar k_{ss}\simeq 10.5$ respectively. Thus we see that, without imposing a constant $\bar k $ as
in~\cite{santos-plos-06,zimmer-pre-05}, $\bar k$ nonetheless tends to increase only slightly,
which nicely agrees with observations of real social 
networks~\cite{barab-collab-02,kossi-watts-06,tom-leslie-evol-net-07}. There is a special case when the
steady-state is all-defect and the simulation is allowed to run for a very long time ($2\times 10^4$ macrosteps);
in this case the link structure never really settles down, since players are unsatisfied, and $\bar k$ may reach a value of about 12 when starting
with $\bar k=10$ and $q=0.8$.
\begin{figure} [!ht]
\begin{center}
	\vspace*{-0.2cm}
\includegraphics[width=14cm] {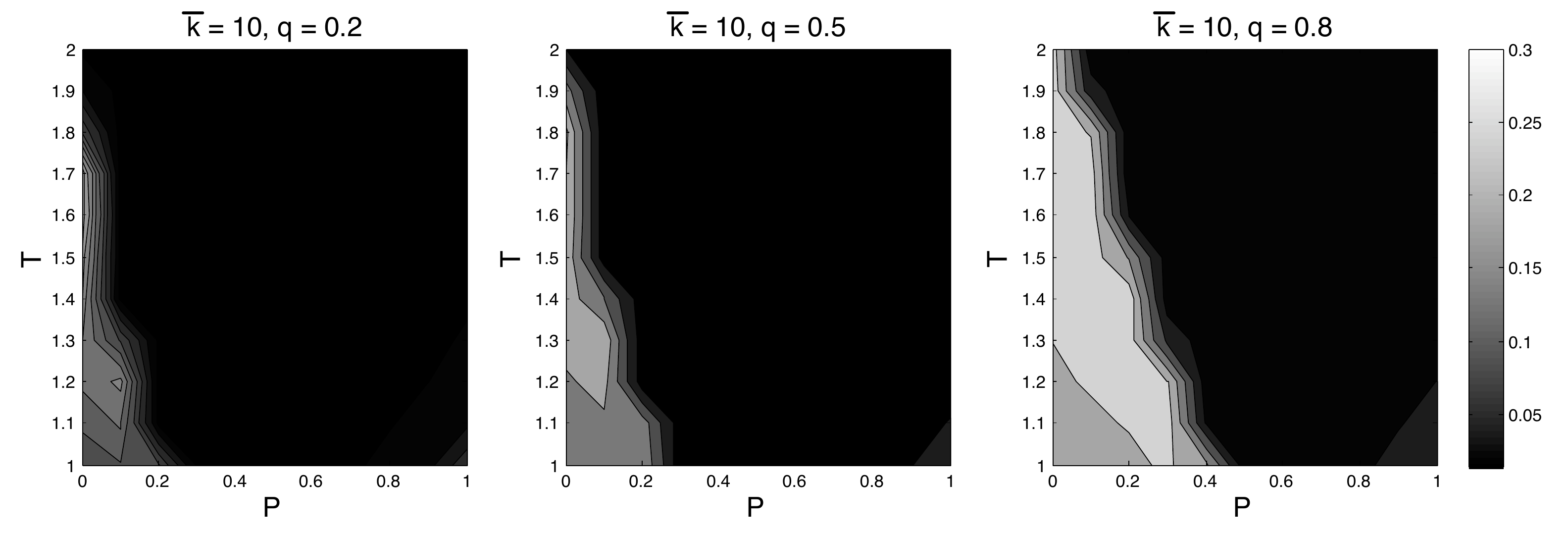} \protect \\	
	\vspace*{-0.2cm}
\caption{Clustering coefficient level for the PD game. Lighter gray means more clustering.\label{pd_cc}}
\end{center}
\end{figure}

Another important global network statistics is the average clustering coefficient $\mathcal C$. The clustering coefficient
$\mathcal C_i$ of a node $i$ is defined as $\mathcal C_i=2E_i/k_i(k_i-1)$, where $E_i$ is the number of edges in  the
neighborhood of $i$. Thus $\mathcal C_i$ measures the amount of ``cliquishness'' of the
neighborhood of node $i$ and it characterizes the extent to which nodes adjacent to node $i$ are
connected to each other. The clustering coefficient of the graph is simply the average over all nodes:
$\mathcal C = \frac{1}{N} \sum_{i=1}^{N} \mathcal C_i$~\cite{newman-03}.
Random graphs are locally homogeneous and for them $\mathcal C$ is simply equal to
the probability of having an edge between any pair of nodes independently. In contrast, real networks have
local structures and thus higher values of $\mathcal C$. Fig.~\ref{pd_cc} gives the average clustering coefficient
$\bar{ \mathcal C}=\frac{1}{50} \sum_{i=1}^{50} \mathcal C$ for
each sampled point in the PD configuration space, where $50$ is the number of network realizations
used for each simulation. It is apparent that the networks self-organize and
acquire local structure in the interesting, cooperative parts of the parameter's space, since the clustering coefficients 
there are higher than that of the random graph
with the same number of edges and nodes, which is $\bar k/N=10/1000=0.01$.
Conversely, where defection predominates
$\mathcal C$ is smaller, witnessing of a lower amount of graph local restructuring. These impressions are confirmed by the
study of the degree distribution functions (see below). The correlation between clustering and cooperation also
holds through increasing values of $q$: $\mathcal C$ tends to increase from left to right in fig.~\ref{pd_cc}, a trend similar
to that observed in the middle row of fig.~\ref{pd_coop} for cooperation. This correlation is maintained also for $\bar k=5$ and
$\bar k=20$ (not shown). 
 \begin{figure} [!ht]
\begin{center}
	\vspace*{-0.2cm}
\includegraphics[width=14cm] {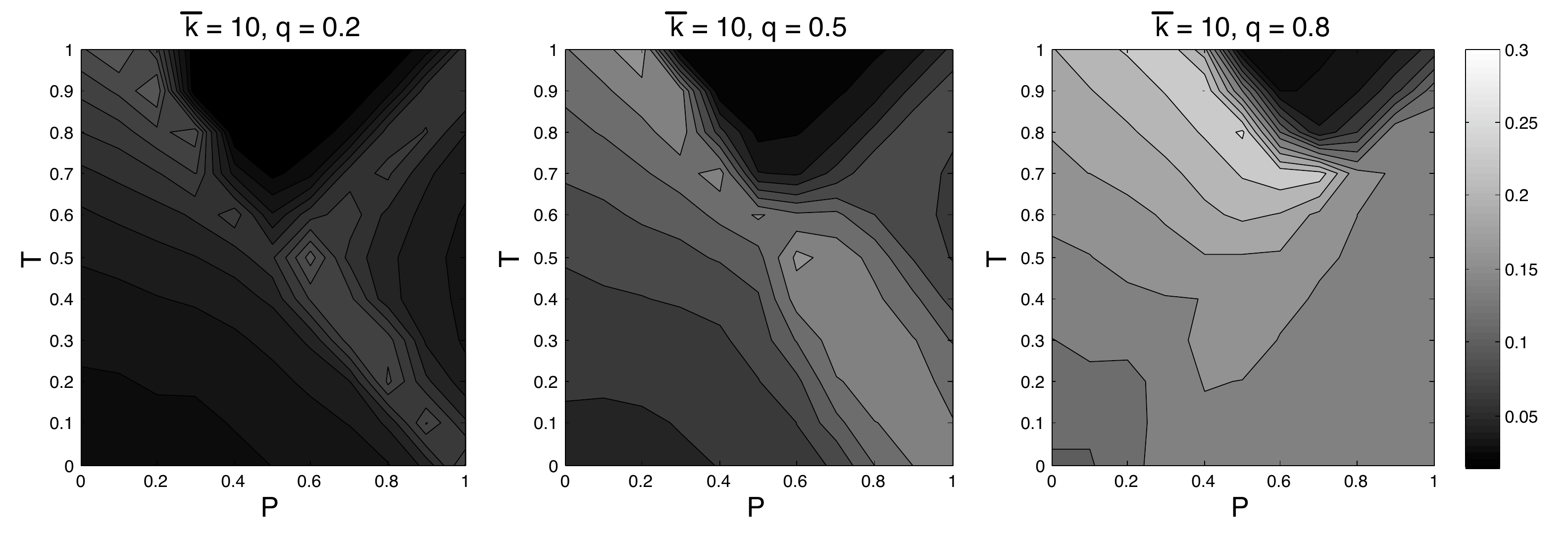} \protect \\
	\vspace*{-0.2cm}	
\caption{Clustering coefficient level for the SH game.\label{sh_cc}}
\end{center}
\end{figure}

As far as the clustering coefficient is concerned, the same qualitative phenomenon is observed for the SH namely,
the graph develops local structures and the more so the higher the value of $q$ for a given $\bar k$
(see fig.~\ref{sh_cc}). Thus, it seems that evolution towards cooperation and coordination passes through a
rearrangement of the neighborhood of the agents with respect to the homogeneous random initial situation,
something that is made possible through the higher probability given to neighbors when
rewiring a link, a stylized manifestation of the commonly occurring social choice of partners. 
\begin{figure} [!ht]
\begin{center}
\begin{tabular}{cc}
	\mbox{\includegraphics[width=6.3cm]{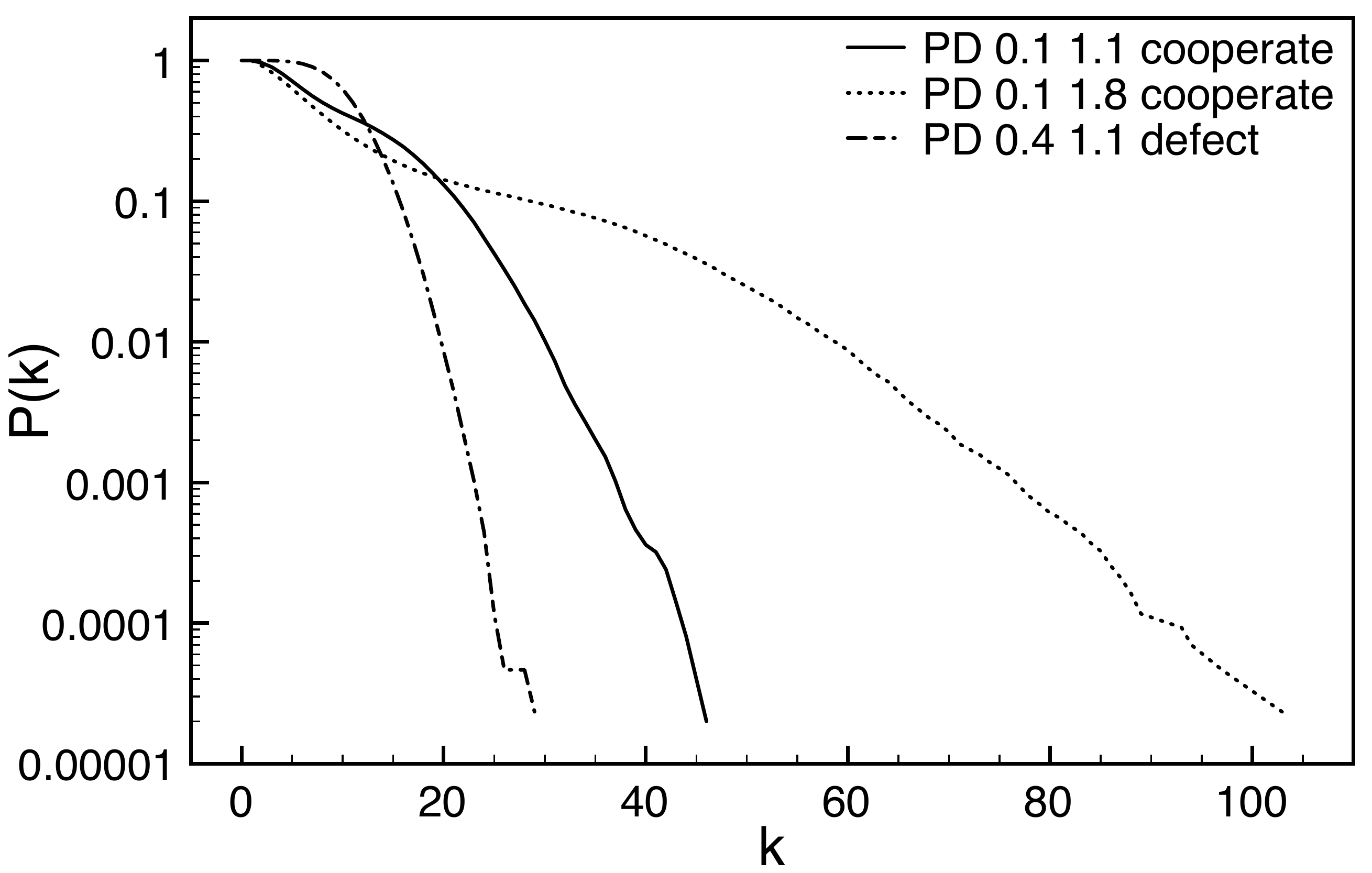}} \protect &
	\mbox{\includegraphics[width=6.3cm]{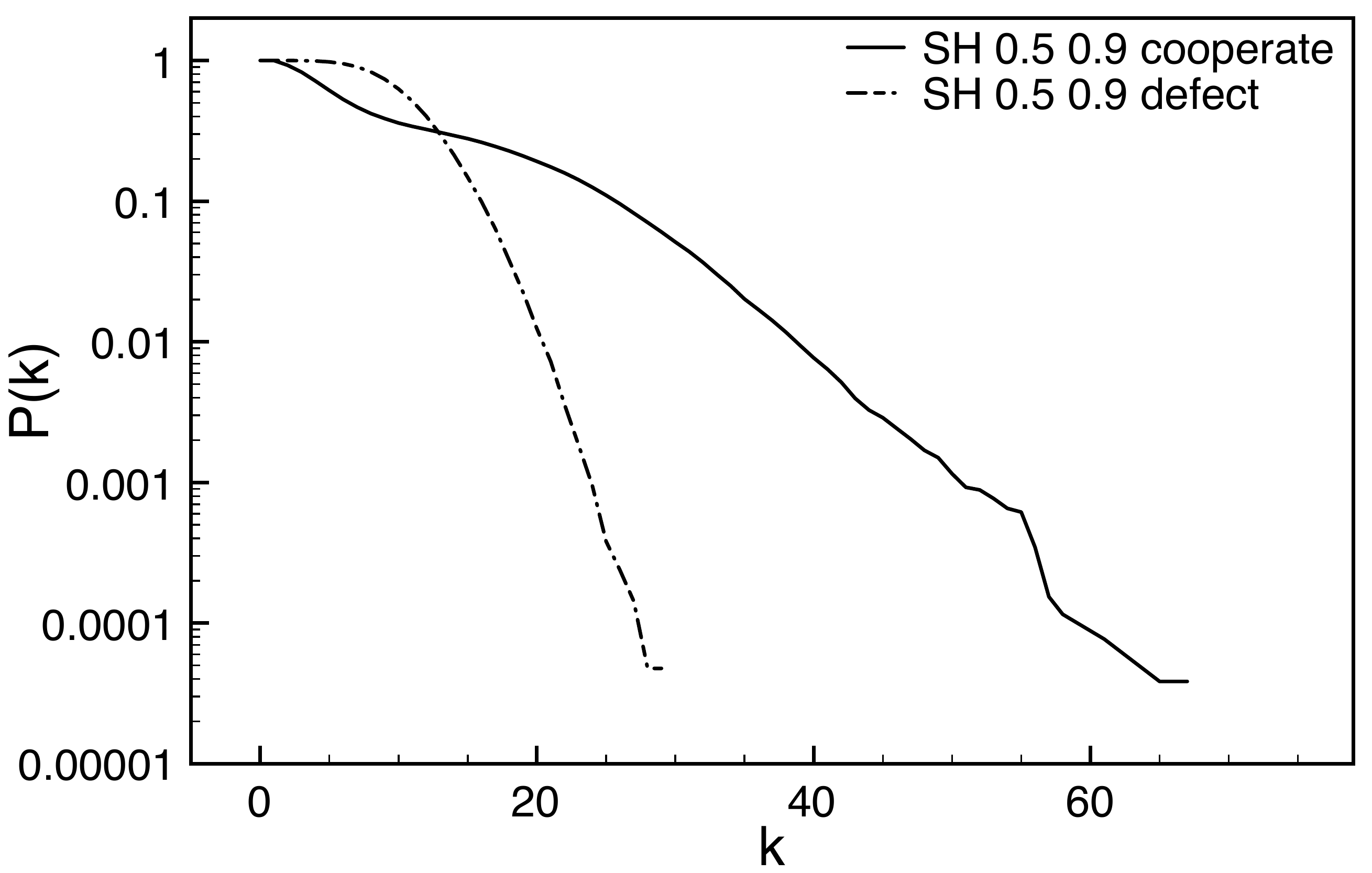}}\protect\\	
	\vspace*{0.5cm}(a)   &  (b) \\
\end{tabular}
\vspace{-0.8cm}
\caption{Cumulative degree distributions. Average values over 50 runs. (a): PD, (b): SH. $q=0.8$, $\bar k=10$. Linear-log scales.
\label{pd_degree}}
\end{center}
\end{figure}

\begin{figure} [!ht]
\begin{center}
	\includegraphics[width=6.3cm]{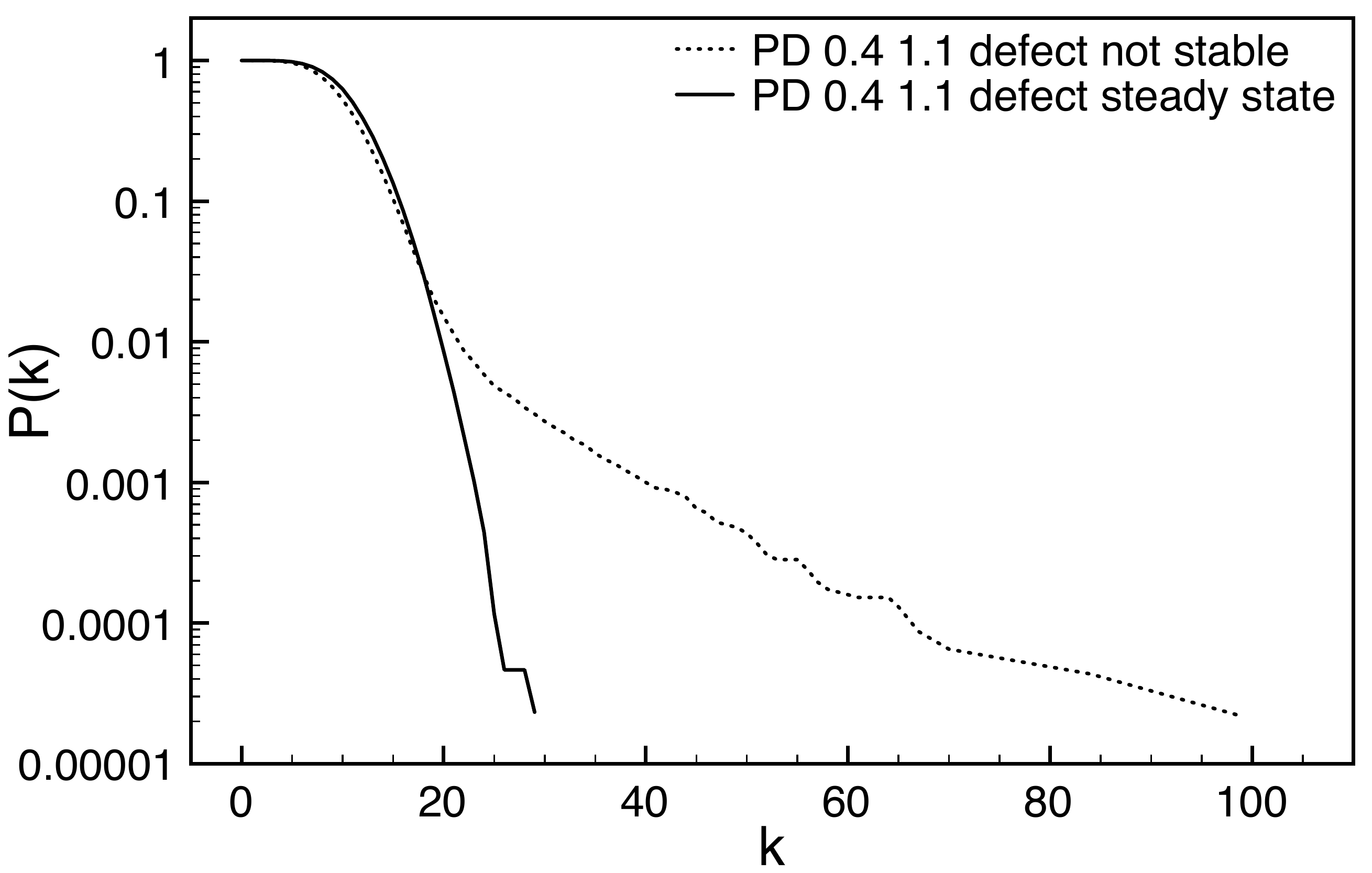} \protect \\
\vspace{-0.2cm}
\caption{Cumulative degree distributions for the PD in case of defection before (dotted line) and after (thick line) reaching a steady-state. Linear-log scales.
\label{pd-trans}}
\end{center}
\end{figure}

The \textit{degree distribution function} (DDF) $p(k)$ of a of a graph represents the probability that a randomly
chosen node has degree $k$~\cite{newman-03}. Random graphs are characterized by DDF of Poissonian
form, while social and technological real networks often show long tails to the right, i.e. there are nodes
that have an unusually large number of neighbors~\cite{newman-03}. In some extreme cases
the DDF has a power-law form $p(k) \propto k^{-\gamma}$; the 
tail is particularly extended and there is no characteristic degree. The \textit{cumulative
degree distribution function} (CDDF) is just the probability that the degree is greater than or equal to $k$
and has the advantage of being less noisy for high degrees.
 Fig.~\ref{pd_degree} (a) shows the CDDFs for the PD for three cases of which two are in the cooperative region and
 the third falls in the defecting region (see fig.~\ref{pd_coop}). The dotted curve refers to a region of the configuration space in
 which there is cooperation in the average but it is more difficult to reach, as the temptation parameter is high
 (T=1.8,P=0.1). The curve has a rather long tail and is thus broad-scale in the sense that there is no
 typical degree for the agents. Therefore, in the corresponding network there are cooperators that are linked
 to many other cooperators. On the other hand, if one considers the dotted-dashed curve, which corresponds
 to a defecting region (T=1.1,P=0.4), it is clear that the distribution is much closer to normal, with a well-defined 
 typical value of the degree.  Finally, the third thick curve, which corresponds to a region where cooperation is more
 easily attained (T=1.1,P=0.1), also shows a rather faster decay of the tail than the dotted line and a narrower
 scale for the degree. Nevertheless, it is right-skewed, indicating that the network is no longer a 
 pure random graph. Since we use linear-log scales, the dotted
 curve has an approximately exponential or slower decay, given that a pure exponential would appear as a straight line in the plot. The tail
 of the thick curve decays faster than an exponential, while the dashed-dotted curve decays even faster.
 Almost the same observations also apply to the SH case, shown in fig.~\ref{pd_degree} (b).
  These are quite typical behaviors and we can conclude that, when cooperation is
 more difficult to reach, agents must better exploit the link-redirection degree of freedom in order for
 cooperators to stick together in sufficient quantities and protect themselves from exploiting defectors during
 the co-evolution. When the 
 situation is either more favorable for cooperation, or defection easily prevails, network rearrangement is less radical. In the limit of long simulation times, the defection case leads to networks that have degree distribution close to Poissonian and are thus almost random. Fig.~\ref{pd-trans} shows such a case for the PD. The dashed curve
 is the CDDF at some intermediate time, when full defection 
 has just been reached but the network is still strongly reorganizing itself. Clearly,
 the distribution has a long tail. However, if
 the simulation is continued until the topology is quite stable at the mesoscopic level, the distribution becomes
close to normal (thick curve).
\begin{figure} [!ht]
\begin{center}
	\includegraphics[width=6.3cm]{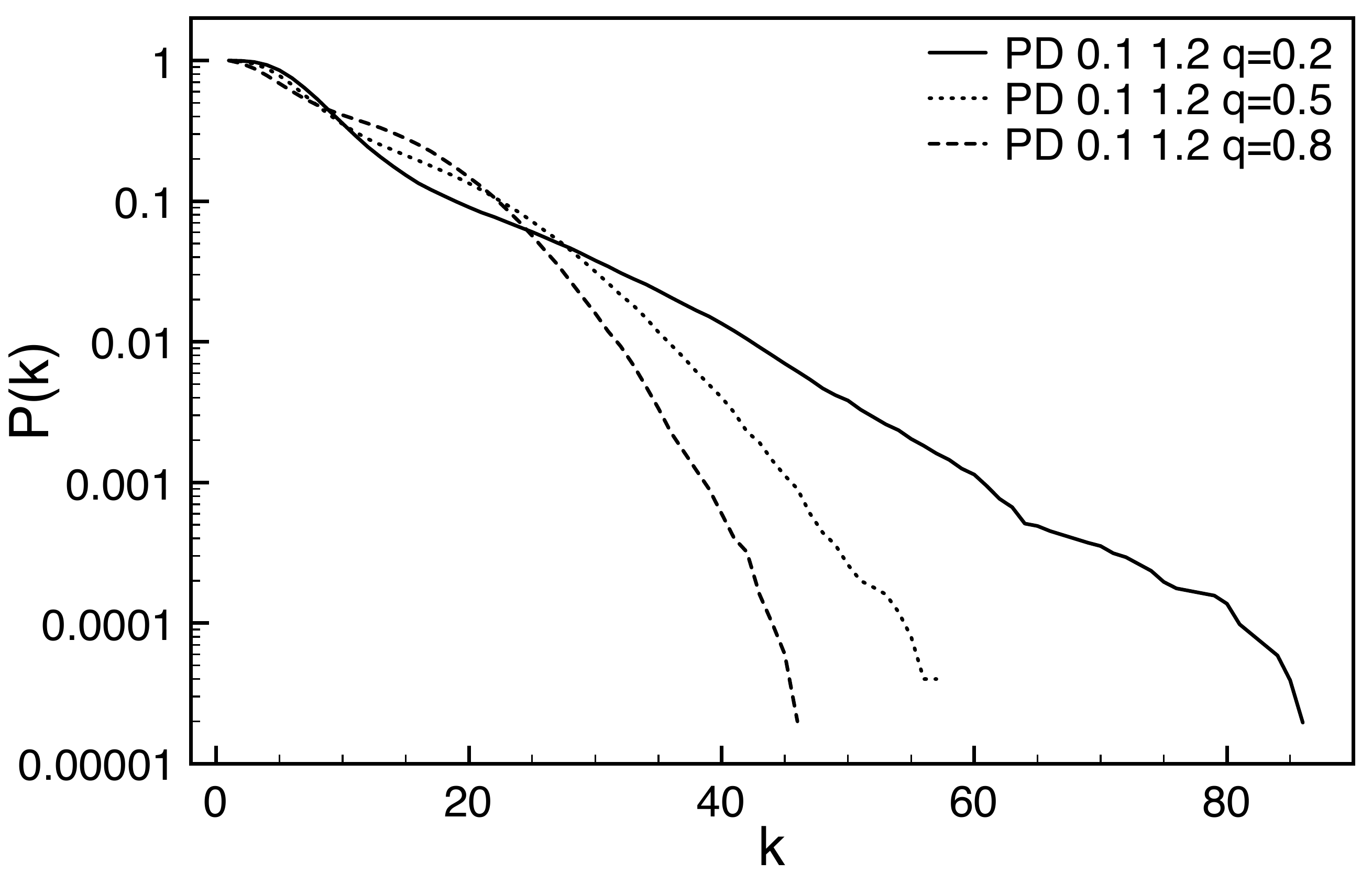} \protect \\
\vspace{-0.2cm}
\caption{Cumulative degree distribution functions for three values of $q$, for the same point in
the PD configuration space in the cooperating region.
\label{pd-cddf-q}}
\end{center}
\end{figure}

Finally, it is interesting to observe the influence of the $q$ parameter on the shape of the degree
distribution functions for cooperating networks. Fig.~\ref{pd-cddf-q} reports average curves for three values of $q$. For high $q$, the cooperating steady-state is reached faster, which gives the network less time to  rearrange its links. For lower values of $q$ the distributions become broader, despite the fact that
rewiring occurs less often, because cooperation in this region is harder to attain and more simulation time
is needed.

\paragraph{Influence of Timing.}
\begin{figure} [!ht]
\begin{center}
	\includegraphics[width=8.3cm]{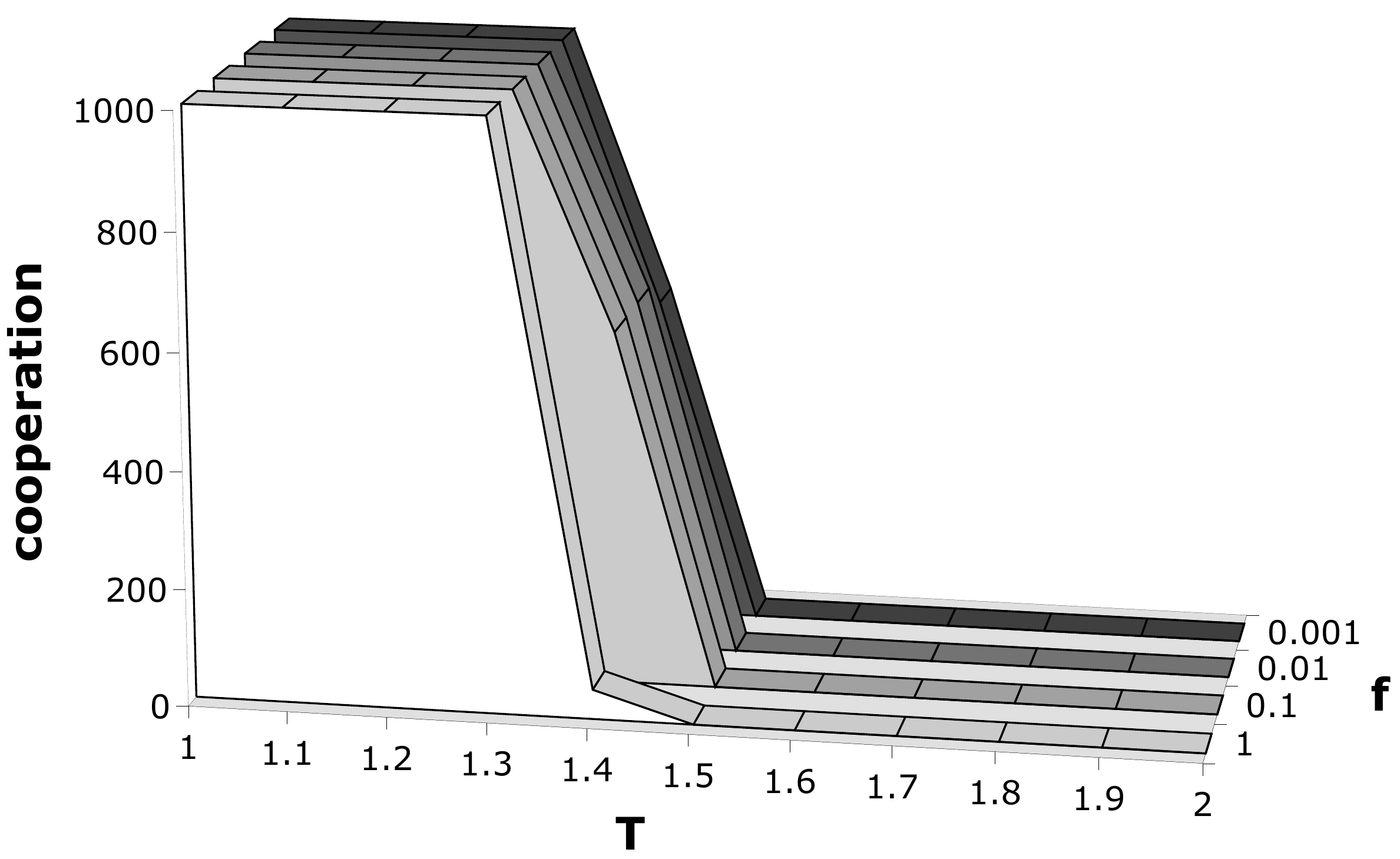} \protect \\
\vspace{-0.2cm}
\caption{Cooperation levels in the PD for $P=0.1$ and $1 \le T \le 2$ as a function of the
synchronicity parameter $f$.
\label{pd_time}}
\end{center}
\end{figure}

Fig.~\ref{pd_time} depicts a particular cut in the configuration space as a function of the
synchronicity parameter $f$. The main remark is that asynchronous updates give similar 
results, in spite of the difference in the number of agents that are activated in a single
microstep. In contrast, fully synchronous update ($f=1$) appears to lead to a slightly less
favorable situation for cooperation. Since fully synchronous update is physically unrealistic and
can give spurious results due to symmetry, we suggest using fully or partially asynchronous update
for agent's simulation of artificial societies.

\subsection{Clusters}

\begin{figure} [!ht]
\begin{center}
	\mbox{\includegraphics[width=5cm]{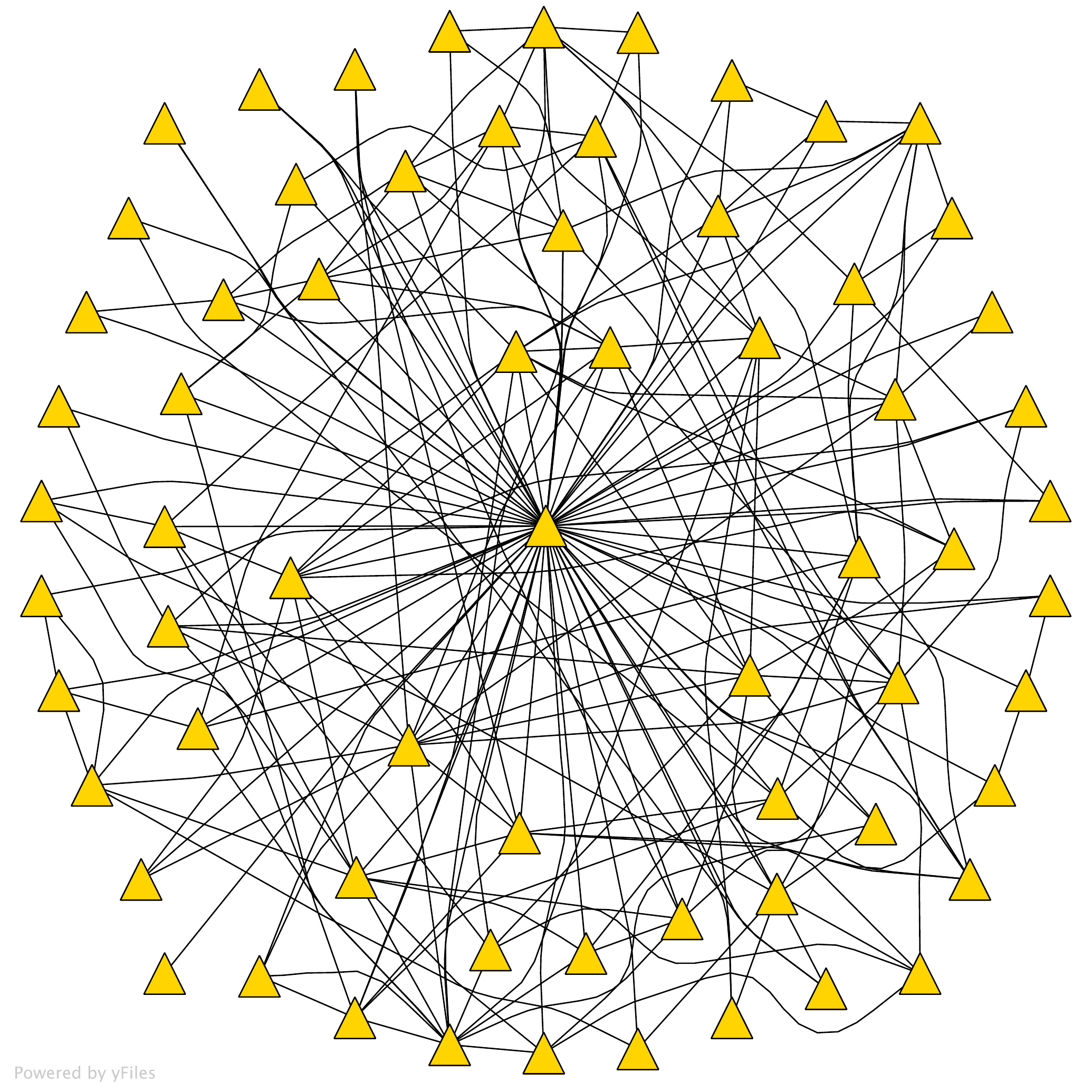}} \protect \\
\vspace{-0.2cm}
\caption{Example of a tightly packed cluster of cooperators for PD networks. $T=1.8, P=0.1$ and $q=0.8$. \label{pd_cluster}}
\end{center}
\end{figure}

We have seen in the previous section that, when cooperation is attained in both games as a
quasi-equilibrium state, the system remains stable through the formation of clusters of
players using the same strategy. In fig.~\ref{pd_cluster}  one such typical cluster corresponding to a situation in which global cooperation has been reached in the PD is shown. Although all links towards the
``exterior'' have been suppressed for clarity, one can clearly see that the central cooperator
is a highly connected node and there are many links also between the other neighbors. Such a
tightly packed structure has emerged to protect cooperators from defectors that, at earlier times, were
trying to link to cooperators to exploit them. These observations explain why the degree  distributions are long-tailed (see previous section), and also the higher values of the clustering coefficient
in this case (see sect.~\ref{topo}).
\label{comm}

\begin{figure} [!ht]
\begin{center}
\begin{tabular}{cc}
	\mbox{\includegraphics[width=5cm]{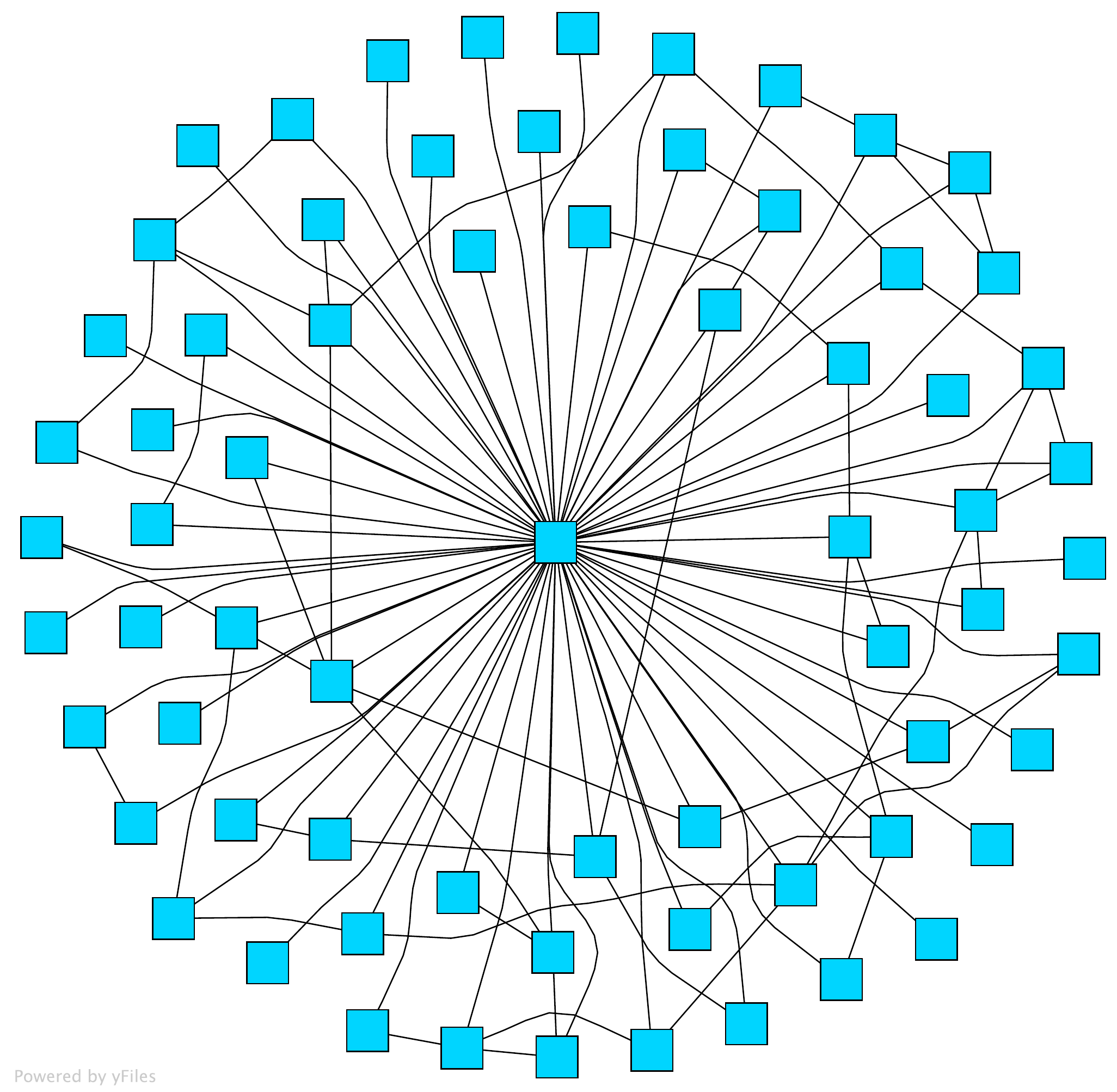}} \protect &
	\mbox{\includegraphics[width=5cm]{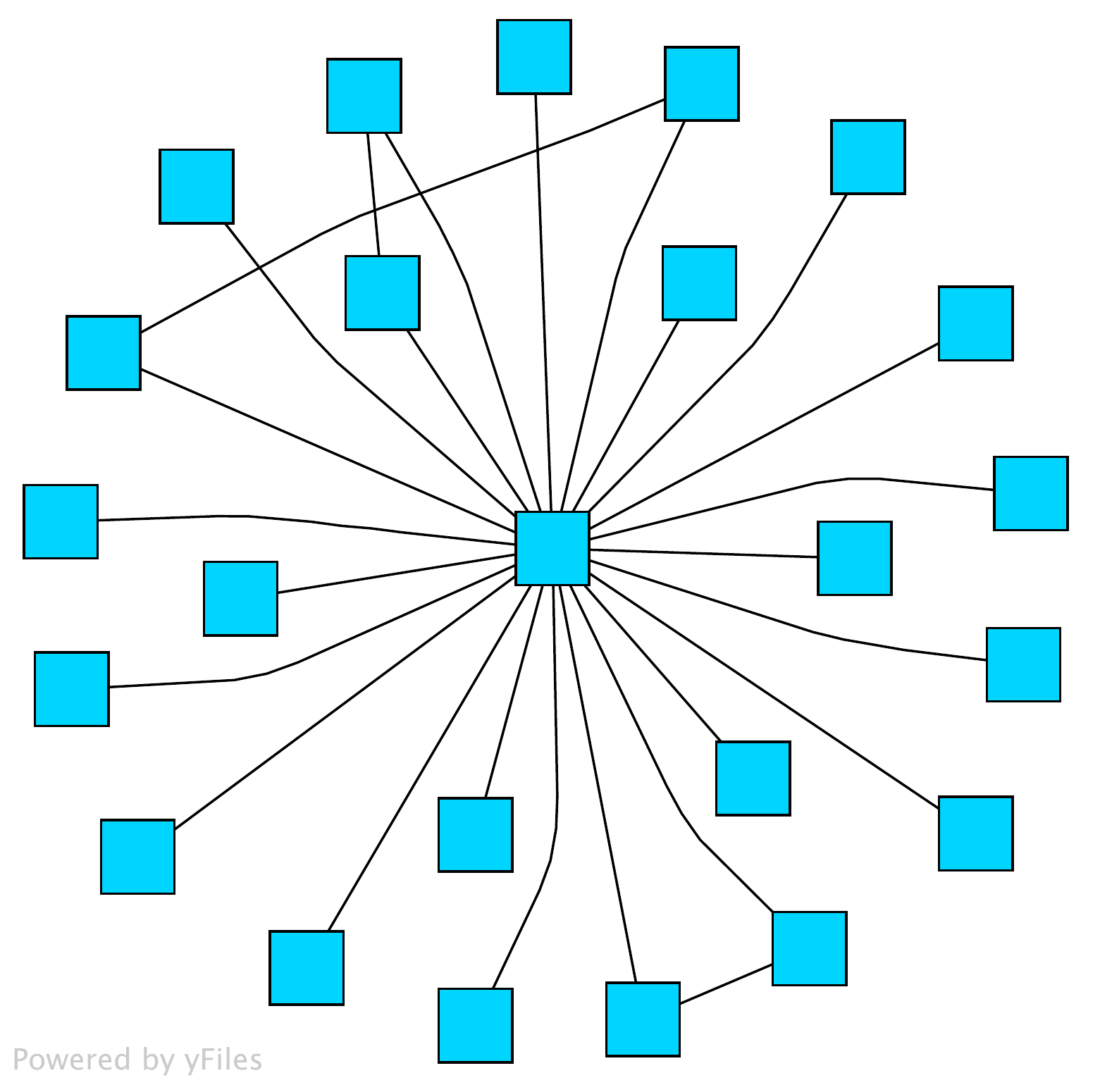}}\protect\\	
	\vspace*{0.5cm}(a)   &  (b) \\
\end{tabular}
\vspace{-0.7cm}
\caption{Example of defector clusters for PD networks, for $T=1.8, P=0.3$ and $q=0.8$.  Clusters like (a) exists only just after  the all-defect state is reached. When a steady-state is reached only clusters like (b) are present in a network of defectors.
\label{pd_cluster2}}
\end{center}
\end{figure}

When the history of the stochastic process is such that defection prevails in the end, the situation
is totally different. Fig.~\ref{pd_cluster2} (a) and (b) show two typical examples of cluster structures
found during  a simulation. Fig.~\ref{pd_cluster2} (a) refers to a stage in which the society is composed
solely by defectors. However, the forces of the links between them are low, and so many defectors
try to dismiss some of their links. This situation lasts for a long simulated time (actually, the system
is never at rest, as far as the links are concerned) but the dense clusters tend to dissolve, giving
rise to structures such as the one shown in fig.~\ref{pd_cluster2} (b). If one looks at the degree
distribution at this stage (fig.~\ref{pd-trans}) it is easy to see that the whole population graph tends to
become random.

The SH case is very similar, which is a relatively surprising result. In fact, when cooperation finally takes
over in regions of the configuration space where defection would have been an almost equally,
likely final state, players are highly clustered and there are many highly connected individuals,
while in less conflicting situations the clusters are less dense and the
degree distribution shows a faster decay of the tail.
On the other hand, when defection is the final quasi-stable state, the population graphs looses
a large part of its structure. Thus, the same
topological mechanisms seem to be responsible for the emergence of cooperation in the PD and
in the SH. The only previous study that investigates the structure of the resulting networks in
a dynamical setting is, to our knowledge, reference~\cite{zimmer-pre-05}, where only the PD is studied. It
is difficult to meaningfully compare our results with theirs as
the model of Zimmermann et al. differs from ours in many ways. They use a deterministic hard-limit 
rule for strategy update which is less smooth than our stochastic local replicator dynamics. Moreover, they study
the PD in a reduced configuration space, only links between defectors can be broken, and links
are rewired at random. They concentrate on the study of the stability of the cooperating steady-states
against perturbations, but do not describe the topological structures of the pseudo-equilibrium states
in detail. Nevertheless, it is worthy of note that the degree distribution functions for cooperators and
defectors follow qualitatively the same trend, i.e.~cooperators networks have distributions 
with fatter tails to the right than defector networks.

\section{Conclusions and Future Work}
\label{conclusions}

Using two well known games that represent conflicting decision situations commonly found in animal and
human societies, we have studied by computer simulation the role of the dynamically networked society's structure in the establishment of
global cooperative and coordinated behaviors, which are desirable outcomes for the society's welfare.  
Starting from randomly connected players which only interact locally in a restricted neighborhood, 
and allowing agents to probabilistically and bilaterally dismiss unprofitable relations and create new ones,
the stochastic dynamics lead to pseudo-equilibria of
either cooperating or defecting agents. With respect to standard replicator dynamics results for
mixing populations, we find that there is a sizable configuration space region in which cooperation may emerge and be stable for the PD, whereas the classical result predicts total defection. For the SH,
where both all-cooperate and all-defect steady-states are theoretically possible, we show that the
basin of attraction for cooperation is enhanced. Thus, the 
possibility of dismissing a relationship and creating a new one does indeed increase the potential
for cooperation and coordination in our artificial society. The self-organizing mechanism consists
in both games in forming dense clusters of cooperators which are more difficult to dissolve by exploiting defectors. 
While the beneficial effect of relational or geographical static population structures on cooperation
 was already known from
previous studies, here we have shown that more realistic dynamic social networks may also allow
cooperation to thrive.
Future work will deal with the stability of the cooperating states against stronger perturbations than
merely the implicit noise of the stochastic dynamics. We also intend to study more fully the structure
of the emerging clusters and their relationships, and we plan to extend the model to other important
paradigmatic games such as Hawks-Doves and coordination games.

\paragraph*{Acknowledgments.}
E. Pestelacci and M. Tomassini are grateful to the Swiss National Science Foundation for financial support under contract number 200021-111816/1. We thank the anonymous reviewers for useful remarks
and suggestions.

\bibliographystyle{plain}

\end{document}